\documentclass[graybox]{svmult}
\usepackage{natbib}
\usepackage{amssymb}
\usepackage{amsmath}
\usepackage{mathptmx}       
\usepackage{helvet}         
\usepackage{courier}        
\usepackage{type1cm}        
\usepackage{makeidx}         
\usepackage{graphicx}        
\usepackage{multicol}        
\usepackage[bottom]{footmisc}
\makeindex             
\begin{document}
\bibpunct{(}{)}{;}{a}{,}{;}

\title*{Dynamics of noncohesive confined granular media}
\author{Nicol\'as Mujica \& Rodrigo Soto}
\institute{Departamento de F\'isica, Facultad de Ciencias F\'isicas y Matem\'aticas, Universidad de Chile}
%
%

\thispagestyle{empty} \maketitle \thispagestyle{empty}
\setcounter{page}{1}

\abstract{Despite the ubiquitousness and technological and scientific importance of granular matter, our understanding is still very poor compared to molecular fluids and solids. Until today, there is no unified description, which indeed seems unreachable. However, it has been proposed that important advances could be attained for noncohesive, hard-sphere like systems, by combining fluid dynamics with phase-field modeling through an appropriate order parameter \citep{RMP06}. Here, we present a review of the dynamics of confined granular matter, for which this systematic approach has proven its value. Motivated by the pioneering work of \citet{Olafsen}, many experimental, theoretical and numerical studies of model confined granular systems have been realized, which have unveiled a very large variety of fundamental phenomena. In this review, we focus on few of these fundamental aspects, namely phase coexistence, effective surface tension, and a detailed description of the liquid state. }

%
%
%

\section{Introduction}
\label{s:intro}
Granular materials are ubiquitous in nature and in everyday life, including industrial applications. Examples are sand dunes, avalanches, granular food of daily usage, transport of grains in food and pharmaceutical industry, transport of grains in the mining industry, granular flow in block caving extraction in a mine, dust disks at the astrophysical scale, and so forth. However, compared to the liquid and vapor states of matter, which flow is well described by the 
hydrodynamic conservation equations, our understanding of the dynamics of granular matter is very limited  \citep{AFP_Book}. It is the athermal nature of the macroscopic constituents of granular matter and their dissipative interactions that generate the main differences with molecular fluids.
The fact is that there is no unified description for the collective motion of granular systems. This is a key problem for the industry, where these materials are the second most important after fluids. Scientifically, granular materials are a prototype of matter far from equilibrium. Advances in our current knowledge of these systems will also strongly impact the general field of non-equilibrium statistical physics. It is known that granular matter can exist in solid, liquid or gaseous states depending on how we force it \citep{RMP96} or the three states can coexist, as during an avalanche. Also they  undergo instabilities and pattern formation \citep{RMP06}. 

Granular systems are simply defined as a collection of macroscopic solid particles for which thermal fluctuations are negligible. Typical sizes range from $1$ $\mu$m to cm scale, or much larger for geophysical and other natural phenomena. 
Additionally, the individual constituents have contact forces that are intrinsically dissipative, given by friction and inelastic collisions. Particles can also interact via electrostatic, van der Waals, magnetic or hydrodynamic forces (e.g. air drag), or even through capillary bridges. Here, for this review, we use the classification given by \citet{AFP_Book}, where particles of size larger than $100$ $\mu$m are considered as granular media, for which in many situations dissipative collisions and frictional contacts are dominant. In this classification, particles of sizes between $1$~$\mu$m and $100$ $\mu$m are considered as powders and below $1$ $\mu$m we refer to colloids. For the former, van der Waals forces, capillary bridge forces and air drag are more relevant than the collisional interactions. For the latter, the thermal nature of the systems is recovered. 

The athermal nature of granular media is a consequence of the macroscopic size of its constituents. Indeed, the potential energy that is required to lift a glass bead of diameter $d = 100$ $\mu$m to a height comparable with this lengthscale is $10^8$ times larger than the thermal energy scale $k_B T$ at ambient temperature. As energy is dissipated at grain-grain and grain-boundary collisions, dynamic states can be obtained injecting energy by different mechanisms, typically via boundaries (vibration or shear) or external fields (e.g. air fluidization) acting on the bulk and, therefore, the detailed balance condition is not fulfilled. However, a global balance between dissipation and energy injection can be made such that the system reaches a steady state. 

\begin{figure}[t!]
\begin{center}
\includegraphics[width=11cm]{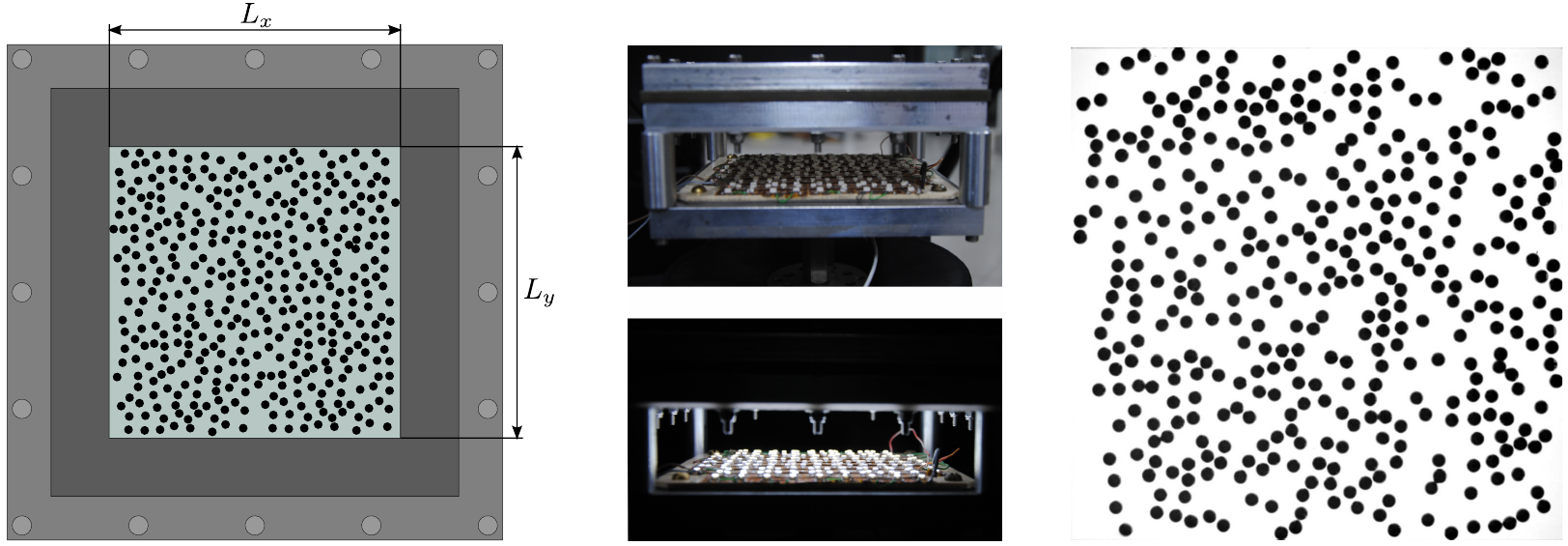}
\caption{Quasi two dimensional (Q2D) geometry using in confined granular experiments. (a) General scheme of the experimental cell, showing the horizontal dimensions and the frame that holds the bottom and top glass plates; (b) Side view pictures of the cell and the base with the illumination LED array off (above) and on (below); (c) Typical image used for particle position detection. From two consecutive images particle velocities are measured as well. Figure from \citet{Neel}. 
\copyright2014 American Physical Society}
\label{Exp_Cell}
\end{center}
\end{figure}

Since some years there has been a large amount of interest in the study of the dynamics of confined, quasi-two-dimensional (Q2D) granular systems, initiated by \citet{Olafsen}, who studied clustering, order and collapse in a vibrated granular monolayer. Since then, simple model experiments and simulations have been developed that allow for a detailed understanding of basic granular physics, with interest for non-equilibrium statistical mechanics and nonlinear physics. More specifically, researchers have focused on the study of vibrated Q2D and Q1D granular systems composed by several hundreds to several thousands spherical particles (of size $\sim 1$ mm in most experiments). The cell's height $H$ is much smaller than the lateral dimensions $L_x$ and $L_y$ (in many cases $L_x = L_y \equiv L)$. 
Energy injection is provided by sinusoidal oscillation of the complete setup, with an imposed vibration $z(t) = A \sin(\omega t)$. In some simulation studies it is provided by a ``thermal" wall, for which particle's velocities are obtained from a Maxwellian flux distribution after a wall collision.  From an experimental point of view, the confined geometry has a great advantage because it allows the observation of both individual trajectories and collective behavior, which enables one to study both the microscopic and macroscopic dynamics.  In general, the focus is on the effective one or two-dimensional system. This dimensional reduction is appropriate as the dynamical time scale of interest is larger than the fast vertical motion, which acts as an energy source. In addition, experiments are also ideal for comparisons with molecular dynamic simulations, which, for example, can explore the vertical motion. Figure \ref{Exp_Cell} presents the  setup used by the authors in various experiments using the Q2D geometry.

In these model systems, individual grains interact solely via collisions. Other forces are negligible except when explicitly introduced, as with electrostatic \citep{Aranson2000,Howell2001} or magnetic interactions \citep{Oyarte,Merminod2014}. The choice of particle size insures that van der Waals forces and  air drag effects are negligible. Also, the use of electrically conducting boundaries (like indium tin oxide, ITO, coated glass plates and metallic walls) minimizes electrostatic interactions. The choice of the particle material allows to avoid magnetic dipolar interactions. Experiments are performed at ambient (low) humidity, so capillary bridge interactions are negligible. Thus, our focus is the study of noncohesive granular matter, which are dominated by dissipative collisions as the main interaction  between particles. 

Although there is no unified description of granular matter, which does seems unreachable for all types of granular systems, \citet{RMP06} proposed that important advances could be attained for dense noncohesive, hard-sphere like systems, by combining fluid dynamics with phase-field modeling through an appropriate order parameter. In this review, we present a set of results concerning the dynamics of confined granular matter for which this systematic approach has proven to work satisfactory. Additionally, these confined systems serve then as a basis for understanding a variety of fundamental phenomena. Here, we will focus on a few of these fundamental aspects, namely phase coexistence, effective surface tension, and detailed description of the liquid state. 

In order to guide the reader to a broader range of phenomena, we provide here a more comprehensive list (still incomplete!) of other fundamental issues that arise in these systems: crystallization \citep{Reis2006}, caging dynamics \citep{Reis2007}, KTHNY-type \citep{Olafsen2005} and surface melting \citep{May}, superheating \citep{Pacheco}, phase separation and coarsening in electrostatically driven systems \citep{Aranson2000,Howell2001} and in dipolar liquids \citep{Oyarte}, the dynamics of  ``polymer-like'' granular chains \citep{BenNaim2001,Safford2009},
the occurrence of giant number fluctuations in nematic monolayers \citep{Narayan2007,Narayan2008,Aranson2008}, segregation in a binary system \citep{rivas2} and related ``granular explosions" \citep{rivas,rivas3}. 
Also, the crucial role of vertical dynamics have been put in evidence 
\citep{rivas,Roeller2011,PerezAngel2011,rivas3,Neel}.

%
%


\section{Granular phase coexistence in confined systems}

\subsection{Clustering in unforced systems}

A remarkable collective effect that is observed in noncohesive granular matter is clustering under certain conditions. A free cooling (unforced) granular gas shows the formation of dense clusters. These clusters are not cohesive and fragment into pieces due to shear stresses but new clusters form dynamically. The mechanism, proposed by \citet{Goldhirsch1993}, is that dense regions created by fluctuations have a higher dissipation rate and therefore cool faster than less dense regions, leading to progressive cooling and clustering. A hydrodynamic analysis shows that this picture is indeed correct, where the clustering is associated to a hydrodynamic instability for long wavelengths \citep{Orza1997}. Beyond a critical system size that depends on density and inelasticity, the compression waves are unconditionally unstable generating the clustering instability and no metastability is observed. In this instability  there is no stable phase separation and no  order parameter presents any non-analytic behavior. Also, there is no evidence of a surface tension that stabilises the interface as it is the case for the transitions described below, which take place in forced confined systems.


\subsection{Liquid-vapor transition}
\begin{figure*}[t!]
\begin{center}
\includegraphics[width=11.5cm]{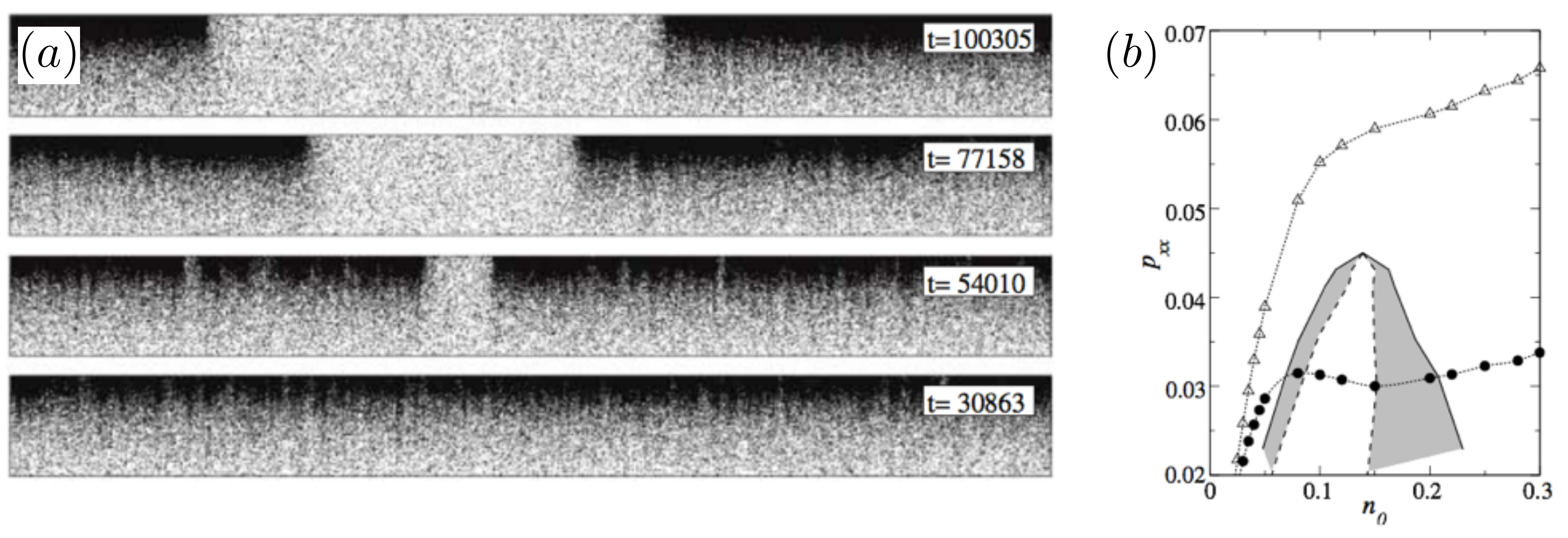}  
\caption{(a): Snapshots of the quasi-1D granular system from simulations, with $L\sim 10000d$ and $H\sim100d$ \citep{Argentina}. Black dots correspond to particles. The bottom wall heats the systems with a ``thermal" boundary condition and periodic boundary conditions are imposed laterally. At the top wall,  collisions are elastic. The systems is initially homogeneous in the longitudinal direction, with a density gradient in the transverse one due to the asymmetry of energy injection. Between the first and second frame a low density ``bubble" nucleates and grows until a stationary coexisting state is attained. (b) Pressure versus filling density computed for a small system, what prevents the instability to develop. At low dissipation, pressure is a monotonic function of density; at larger dissipation, a negative compressibility regions develops. The continuous line, dashed line and the gray region in between correspond to the coexistence line, spinodal line and the bistability region respectively. Copyright \copyright  2002, American Physical Society}
\label{fig1}
\end{center}
\end{figure*}

Forced granular systems that can reach fluid-like steady states also show clustering and phase separation. Some mechanisms are shared with the unforced case  but the picture is quite different.
\citet{Argentina} introduced a simple model were through molecular dynamic simulations it was shown that a heated granular system in 2D can undergo a phase separation, analogous to the spinodal decomposition of the gas-liquid
transition in the van der Waals model (see Fig. \ref{fig1}). This phase separation is triggered by a negative compressibility. The control parameter is the restitution coefficient; at low dissipation, the effective pressure is a monotonically increasing function of density, whereas for larger dissipation a decreasing pressure as function of density is observed. At the onset of this phase transition, the system reveals rich dynamic
behavior characterized by the appearance, coalescence, and disappearance of clusters. Clusters undergo a coarsening process that can not be explained by a standard diffusion mechanism. Instead, it was shown that the system sustains waves, which in turn cause clusters to interact and coarsen much more effectively than through pure diffusion.

Considering the observed propagating waves, which are created through density fluctuations, an order parameter was defined as the density difference with respect to a critical value. Using symmetry arguments, \citet{Argentina} deduce an effective non-linear wave equation, which includes an interface energy term and dissipation. This modified wave equation was named the van der Waals normal form. This theoretical model recovers well the observations obtained by molecular dynamic simulations. In particular, the phase diagram can be computed, displaying a critical point, the spinodal and coexistence curves, with the bistability region in between. \citet{Cartes} deduced the same van der Waals normal form starting from a one-dimensional granular hydrodynamical model. The equation of state, postulated as the one of a hard-sphere disk system in 2D, can present a negative compressibility region as a consequence of the granular temperature being a decreasing function of density. 

This liquid-gas phase separation was also studied experimentally and numerically by \citet{Roeller2011} in a Q2D configuration. Submillimeter glass spheres are vertically shaken in a box with large horizontal dimensions, $L\approx210d$ and much smaller height $H\sim 10d$. For a given particle volume fraction, the system displays an homogeneous liquid state at low vibration amplitudes and a coherent gas state at high amplitudes. The liquid-gas phase separation is observed at intermediate amplitudes; the transitions between the homogeneous states and the phase separation can be obtained by either varying the driving amplitude or the particle volume fraction, with a larger coexistence region for larger $H$. Their event-driven simulations suggest that the phase separation is driven by spinodal decomposition and pressure measurements for small systems display a negative compressibility region. 

\subsection{Solid-fluid transitions}

\subsubsection{Clustering and collapse of submonolayers for low vibration amplitudes}
 
\begin{figure*}[t!]
\begin{center}
\includegraphics[height=3.5cm]{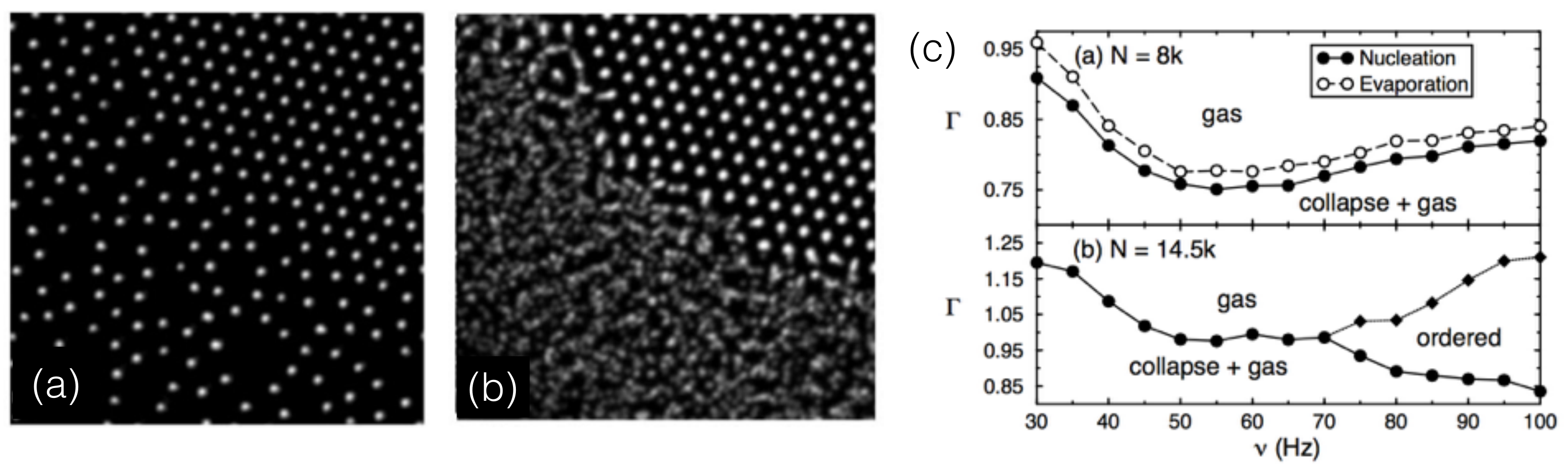}  
\caption{Instantaneous (a) and average (b) photographs of part of the 2D granular system at $\nu = \omega/2\pi = 70$ Hz and $\Gamma = 0.76$. The less dense, gas phase is at the bottom-left half of the image and the more dense, solid cluster at the other half, which is more evident in the average image. Particles in the solid cluster have much lower velocities. (c) Phase diagrams obtained at low (top) and high filling density (bottom). The control parameters are the vibration frequency $\nu$ and normalized acceleration $\Gamma$. Figures from \citet{Olafsen}. Copyright \copyright  1998, American Physical Society}
\label{figOlafsen}
\end{center}
\noindent
\end{figure*}

In their pioneering work, \citet{Olafsen} demonstrated that a 2D granular gas can undergo a clustering and ordering transition by adjusting very carefully the forcing amplitude. A submonolayer of stain-less steel particles is placed on a sinusoidally vibrating plate. At fixed frequency, by decreasing the normalized acceleration, $\Gamma = A\omega^2/g$, to values $<1$, a gas state will phase separate into a gas phase and an hexagonally packed solid cluster. Previous to this phase separation, small density fluctuations are observed, which also present lower velocities. These dense clusters serve as nucleation points for a collapse, from which the hexagonally ordered cluster will grow until it reaches a stable size, in mechanical equilibrium with the gas, low density phase. This dense phase is what they refer as a collapsed state, mainly because of its low kinetic energy. An example of the coexistence of the two phases is shown in Figs. \ref{figOlafsen}(a) and  \ref{figOlafsen}(b). At higher filling densities, the system shows the presence of an hexagonally ordered phase that extends over the entire cell, with no gas phase in equilibrium. In this state, particles fluctuate around average positions but with much larger kinetic energy than the so called collapsed state. Two phase diagrams, for low and high filling densities, are presented in Fig. \ref{figOlafsen}(c), with vibration frequency $\nu=\omega/2\pi$ and acceleration $\Gamma$ as control parameters.  At low densities, the transitions present hysteresis. 

\citet{Olafsen} also present a detailed statistical analysis of the gas positional structure. The experimentally computed pair correlation function was shown to be quantitatively similar to the one obtained with a  Monte Carlo calculation of a 2D gas of elastic hard disks in equilibrium for a the same filling density of the experiment, with no adjustable parameters, showing that the structure of the 2D granular gas is dominated by excluded volume effects. Additionally, spatial correlations were shown to increase when the system is cooled to a lower $\Gamma$, closer to the transition line.

\citet{Losert1999} studied a reverse phase transition in the same experimental configuration. Instead of cooling the granular gas to a collapse, they started with an amorphous submonolayer of particles at rest with $\Gamma \lesssim 1$, which in our current context can be considered as the condensed, solid phase. In a region of the frequency-amplitude parameter space, when an external perturbation is applied to a group of grains, these start to bounce and trigger a transition to a gas state that involves all the particles. The authors focus on the front propagation of this gas state that grows within the also stable amorphous state. At low coverages, they show that these two metastable dynamical states can simultaneously coexist, inferred from observations of particle trajectories; some correspond to stationary particles and some to moving, bouncing particles.

The coexistence of two metastable states was confirmed recently by \citet{PerezAngel2011}. Using 3D molecular  dynamic simulations of a granular system that is vertically vibrated, they demonstrated the coexistence of two type of dynamical modes. Some particles rebound on the vibrating plate (``bouncers"), conforming collectively the gas state, while others remain always in contact with the plate and roll on it (``rollers"). Their respective population depends on the initial conditions that are set. The bouncers in fact bounce with the same period of the vibrating plate. Clusters of rollers and bouncers can form depending on which has the largest initial population; in both cases, clusters form due to the external pressure exerted by the other species. When the largest initial population corresponds to bouncers, the clusters of rollers that are formed end up with an hexagonal order, as in the original experiments by \citet{Olafsen}.

In the Q2D geometry, individual grains can reach a simple limit cycle where they bounce back and forth with the top and bottom walls with the same period of the vibrating cell. When the particle concentration is large, this limit cycle is difficult to reach because collisions with other grains are more probable. The fluid state and the synchronized limit cycle coexist. The later state is absorbing--defined as one from which the system cannot escape--with no residual horizontal dynamics. The former is metastable and after long, but finite, times the system transits from the fluid to the absorbing state. Only large perturbations, that appear in experiments due to imperfections are able to refluidize the system \citep{Neel}.

\subsubsection{Ordering for large vibration amplitudes}

When a Q2D monolayer of mono-disperse noncohesive spherical particles is vertically vibrated, a solid-liquid-like transition occurs when the driving amplitude or filling density exceeds a critical value \citep{Prevost,Urbach3}. This is very counter-intuitive, because the system suffers a transition from a liquid phase to a coexisting solid-liquid state when the vibration amplitude is increased for a fixed density. Thus, solidification occurs while the injected kinetic energy is increased above a certain threshold. Here, we stress that the physical mechanism underlying particle clustering relies on the strong interactions mediated by grain dissipative collisions in a confined geometry, rather than on grain-grain cohesive forces in a non-confined system \citep{Royer}. Varying the filling fraction and box height, different possible solid phases are observed~\citep{Urbach3}. The effect of inelasticity was studied experimentally and numerically by \citet{Reyes2008}; two sets of experiments were performed, with stain-less steel and brass particles, which are less and more dissipative respectively. They showed that inelasticity has a strong effect on the phase diagram, pushing the transition lines to higher critical accelerations for the more dissipative, brass particles. They also demonstrated that the solid cluster can melt again for higher vibration amplitudes or lager inelasticity.  

This phase separation is triggered by the negative compressibility of the effective 2D equation of state and the associated transient dynamics is governed by waves. Indeed, following the pressure measurement setup proposed by \citet{Geminard2004}, \citet{Clerc} demonstrate that the pressure versus density curve reaches a {\it plateau} precisely at the solid-liquid phase coexistence. They also performed 2D and 3D event driven simulations from which pressure measurements confirm the van der Waals scenario for this transition. Using a quasi-one-dimensional setup they demonstrate trough a coarse-graining procedure that waves are induced by fluctuations, but are only observable in the longitudinal momentum field. In this case, solid clusters interact and merge, also much more effectively than a pure diffusive process, as for the liquid-gas van der Waals transition \citep{Argentina,Cartes}.

The effect of forcing and dissipation were studied numerically by \citet{Lobkovsky2009}. They report results from two numerical studies, namely soft-sphere molecular dynamics using a vibrating wall as the forcing mechanism and hard-sphere event driven simulations with random energy injection. For both types of forcing, inelasticity suppresses the transition to the solid-liquid coexistence, shrinking the phase coexistence region for lager dissipation. However, for the random forcing there is no clear phase separation as there is for the vibrating case; indeed, these authors conjecture the existence of an effective surface tension for the latter, although no measurement was proposed. They relate this surface tension to the observation that there is a strong phase-dependence of energy injection, which is minimum at the solid cluster. 

\begin{figure*}[t!]
\begin{center}
\includegraphics[height=5.5cm]{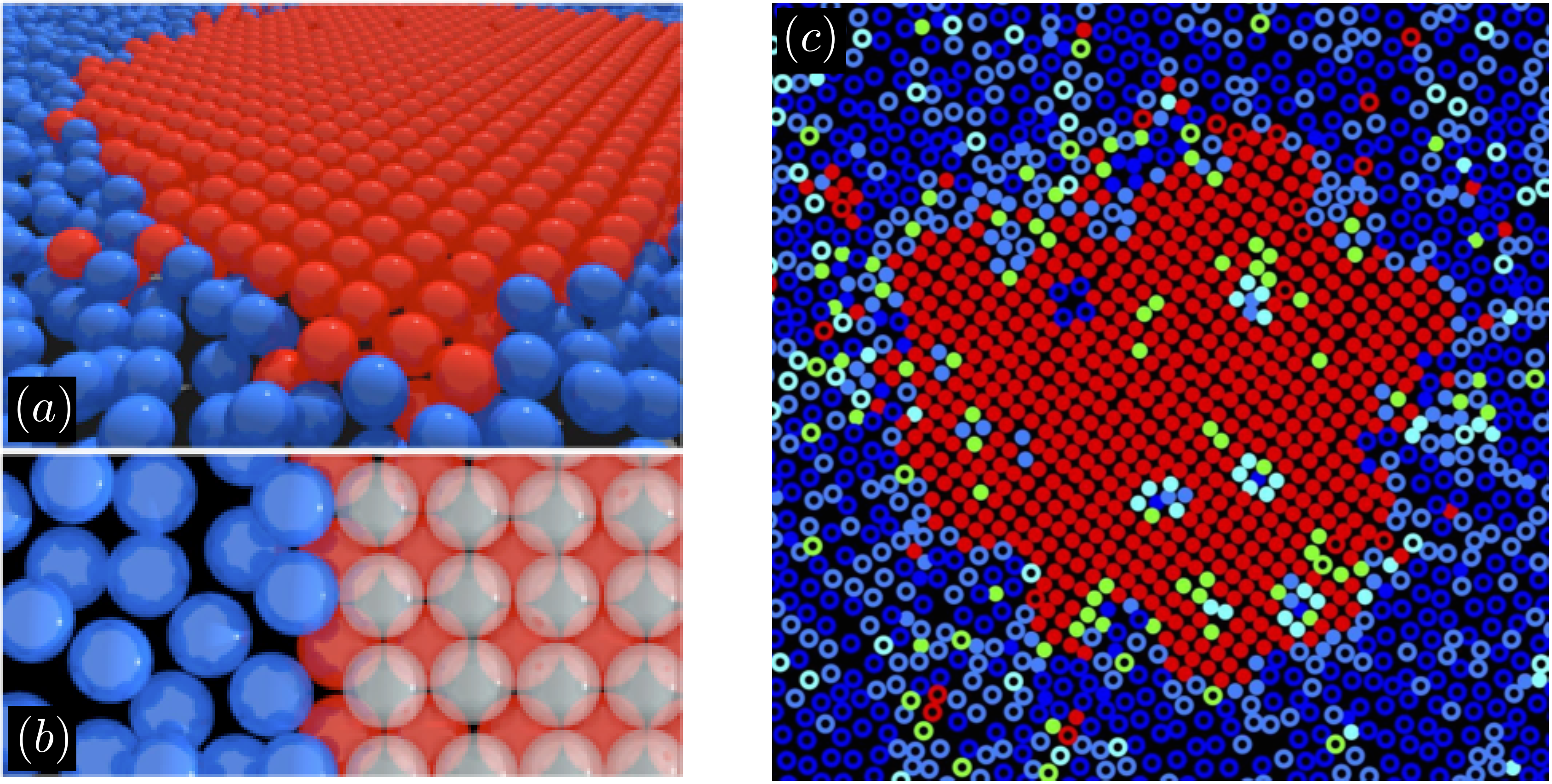} 
\caption{(a-b) Visualization of solid and liquid phases (red-white and blue respectively) from molecular dynamic simulations. The bottom panel shows that the solid cluster is composed by two interlaced layers (white on top and red below). Figure from \citet{Prevost}. Copyright \copyright  2004, American Physical Society. (c) Experimental colormap of the bond-orientational order parameter $Q_4$, from \citet{Castillo}. Red color indicates particles with local order close to a square lattice; blue, those with order far from the square lattice. Solid symbols ($\bullet$) represent dense regions, measured through de Voronoi area; open symbols ($\circ$) correspond to less dense regions. \copyright  2012, American Physical Society}
\label{fig_Q4}
\end{center}
\noindent
\end{figure*}

\begin{figure*}[t!]
\begin{center}
\includegraphics[height=5.1cm]{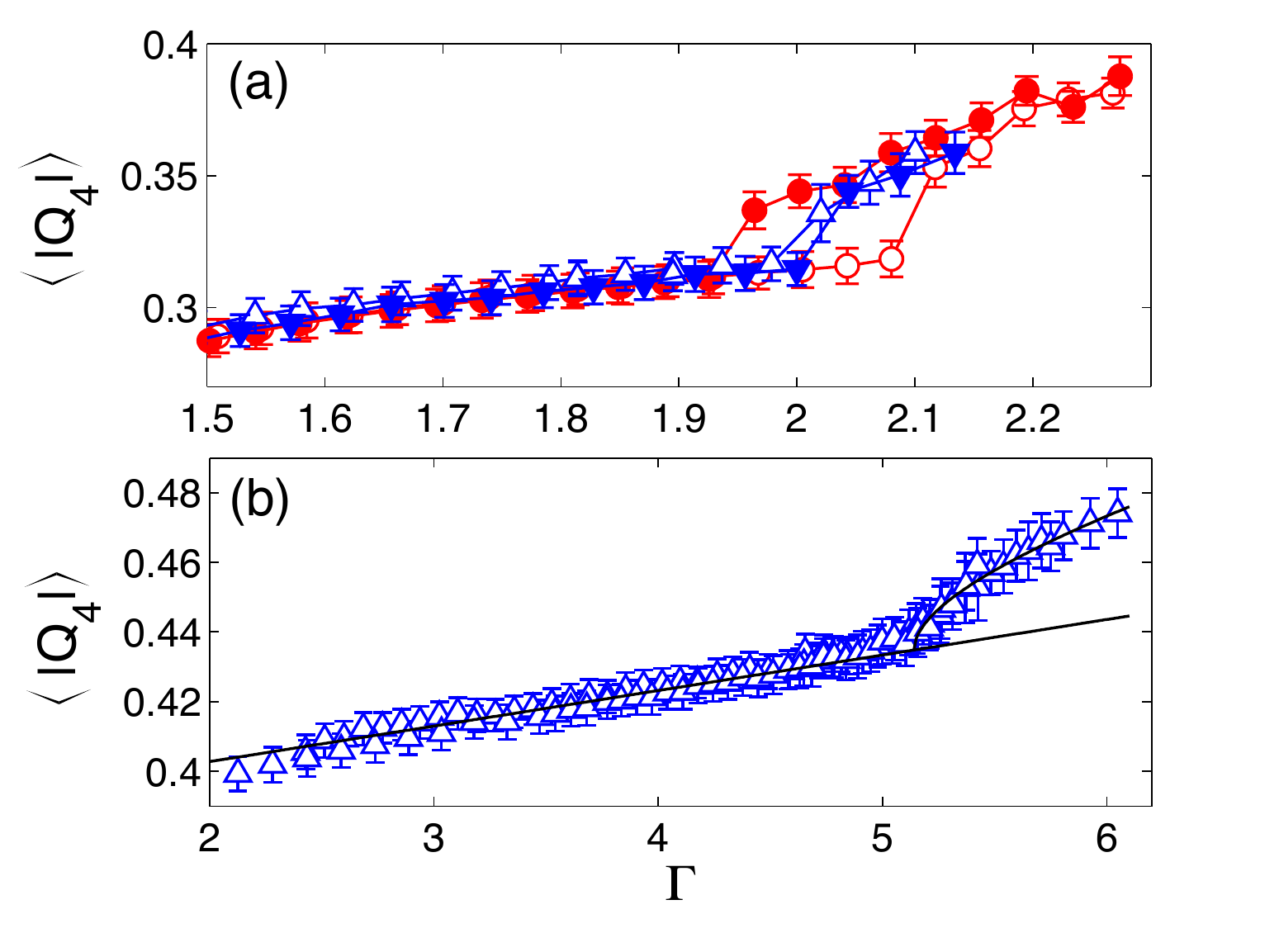} 
\caption{Average global 4-fold bond-orientational order parameter $\langle |Q_4| \rangle$ versus  $\Gamma$ for C1 (a) and C2 (b). Open (solid) symbols represent data obtained for increasing (decreasing) $\Gamma$ ramps, with the following rates: $\Delta\Gamma/\Delta t \approx 0.005$ min$^{-1}$ ({\color{blue} $\triangle$, $\blacktriangledown$}) and $\Delta\Gamma/\Delta t \approx 0.02$ min$^{-1}$ ({\color{red} $\circ$, $\bullet$}). Continuous lines in (b) correspond to fits of the linear trend $Q_4^L = a\Gamma +b$ for $2.5<\Gamma<5$, with $a=0.011\pm0.001$ and $b=0.380\pm0.002$, and a supercritical-like behavior $ \langle |Q_4| \rangle=Q_4^L + c(\Gamma-\Gamma_c)^\beta$, with $\beta=1/2$, observed for $\Gamma \gtrsim 5$. Figure from \citet{Castillo}. \copyright  2012, American Physical Society.}
\label{fig_Q4vsGamma}
\end{center}
\noindent
\end{figure*}

To correctly describe the solid-liquid transition a new order parameter was required. \citet{Castillo} introduced an orientational order parameter (see Fig. \ref{fig_Q4}), which shows that the transition can be either of first or second order type (see Fig. \ref{fig_Q4vsGamma}). Depending on the cell's height and filling density, the transition can be abrupt or continuous. The two configurations that lead to each type of transition referred to as C1 and C2. The 4-fold bond-orientational order parameter per particle is defined
\begin{equation}
Q_4^j = \frac{1}{N_{j}} \sum_{s = 1}^{N_{j}} e^{4i\alpha_{s}^j},
\end{equation}
where $N_j$ is the number of nearest neighbors of particle $j$ and $\alpha_s^j$ is the angle between the neighbor $s$ of particle $j$ and the $x$ axis.  For a particle in a square lattice, $|Q_4^j | = 1$, and the complex phase measures  the square lattice orientation respect to the $x$ axis. Figures \ref{fig_Q4}(a) and \ref{fig_Q4}(b) present results from molecular dynamic simulations obtained by  \citet{Prevost}. Here, the double layered solid cluster can be clearly observed. Figure \ref{fig_Q4}(c) presents a representation of local order $|Q_4^j|$ in the experimental system \citep{Castillo}. Only part of the system is shown, approximately $L_x/3 \times L_y/3$. From this figure it is evident that the denser regions are where the system is more ordered. 

For the second order-type transition, criticality has been reported and five independent critical exponents have been measured \citep{Castillo}. These critical exponents are associated to different properties of $Q_4$: the correlation length, relaxation time, vanishing wavenumber limit (static susceptibility), the hydrodynamic regime of the pair correlation function, and the amplitude of the order parameter. The results are consistent with model C of dynamical critical phenomena \citep{hohenberg}, valid for a non-conserved critical order parameter (bond-orientation order) coupled to a conserved field (density). An example of the observed critical behavior is presented in Fig.  \ref{fig_Q4vsGamma}(b), in this case for the amplitude of the order parameter. 

More recently, \citet{CastilloPRE} tested the universality of this transition by changing the system's dissipation parameters. The top and bottom glass plates have an ITO coating on their inner surfaces that minimizes electrostatic charges. Thus, two different thicknesses of this coating were used, with a thicker (thinner) coating having  lower (larger) effective dissipation parameters. The difference between the dissipation parameters is in fact deduced from the difference in the critical accelerations, with a ratio of about $0.87$.  Within experimental errors, these authors  demonstrate the agreement of all five critical exponents. This is an exigent test for the universality of the transition. Thus, while dissipation is strictly necessary to form the crystal, the path the system undergoes toward the phase separation is part of a well-defined universality class. 

\subsection{Effective surface tension}
\label{subs:surfacetension}

Although our focus is on phase coexistence, we must note that clustering may also occur without phase separation, through capillary effects induced by an effective surface tension. The sole existence of property is  counterintuitive because, due to the dominance of collisional, dissipative interactions, for a long time granular matter has been considered to have no surface tension at all. However, experiments performed with tapped powders show a Rayleigh-Taylor-like instability \citep{Duran}. Also, interface fluctuations and droplet formation in granular streams have been reported both in air \citep{Amarouchene} and vacuum \citep{Royer}, and their origin seems to be the surface-tension-driven Plateau-Rayleigh instability. 

In all these systems, effective surface tensions have been estimated, being as low as $0.1$ $\mu$N/m for the granular free-falling jet in vacuum \citep{Royer}. This is about $10^6$ times lower than the surface tension of a pure air-water interface ($\gamma_{\rm w}\approx 70$~mN/m). Such low surface tension is induced by nanoNewton level cohesion forces, which result from a combination of van der Waals interactions and capillary bridges. In other cases, for the so called strongly forced systems for which momentum transfer dominates other forces, granular matter has shown to behave as a zero-surface-tension liquid, as for particle sheets created by a granular jet impacting a target \citep{Cheng2007} and fingering in a granular Hele-Shaw system \citep{Cheng2008}. In some cases, an hydrodynamic derivation taking the zero-surface-tension limit succeeds in describing the observations \citep{Cheng2007,Cheng2008} but in others a finite surface tension is needed \citep{Ulrich2012}. The key point is to understand how capillary-like features can emerge out of collections of particles that are completely noncohesive or that have very low interaction forces. 

The systems presented in the two previous subsections do present features that are related to the existence of an interface energy, for which surface tensions can be defined and measured. Both system classify as strongly forced systems for which momentum transfer trough collisions dominates other forces. Nonetheless, low surface tensions--of the order of $1$ mN/m--have also been measured in these strongly agitated noncohesive granular systems that phase separate \citep{Clewett2012,Luu}. To our knowledge, these two systems are the only examples that have shown the emergence of an effective surface tension induced by dissipative collisions rather than by low cohesive forces \citep{Royer} or by air drag \citep{Duran,Amarouchene}. 

Using experiments and soft-sphere molecular dynamic simulations \citet{Clewett2012} show that at the liquid-gas coexistence, stable liquid droplets have in average a circular shape. A pressure difference between the liquid drop and the surrounding gas was measured in the molecular dynamic simulations. It was shown to be consistent with a Laplace law, $\Delta P \sim 1/R_0$, with $R_0$ the average radius. This method gives a surface tension $\gamma = 1.0 \pm 0.1$ mN/m for the particular parameters that were used (geometry, filling fraction, forcing frequency and amplitude). Computations of the pressure tensor at the interface showed that the surface tension is mainly given from an anisotropy in the kinetic energy part of the pressure tensor. For a circular drop, surface tension is known to have a mechanical definition, namely
\begin{equation}
\gamma = R_0 \int_{0}^\infty \frac{p_N(r) -p_T(r)}{r} dr, 
\end{equation}
where $p_N$ $(p_T)$ is the normal (tangential) component of the pressure tensor. Evaluating this integral over several drop sizes, they obtained $\gamma = 0.94 \pm 0.05$ mN/m, consistent with the value computed from the Laplace pressure difference.

At the solid-liquid coexistence, \citet{Luu} show that a solid cluster, in average, resembles a drop, with a striking circular shape. The coarse-grained solid-liquid interface fluctuations turn out to be well described by the capillary wave theory, which allows to define an effective surface tension and interface mobility for this noncohesive granular system. More specifically, by computing the Fourier modes of the interface radial fluctuations $\delta R(\theta,t) = R(\theta,t) - R_0$ and kinetic energy $K$, they demonstrated that the system obeys a capillary spectrum  
\begin{equation}
\langle | \widetilde {\delta R}_{|m|\geqslant 2} |^2\rangle = \frac{\langle K_{|m|\geqslant 2} \rangle R_0}{\pi \gamma_{\rm 2D} (m^2-1)},
\end{equation}
where $m$ is the Fourier mode number. For the continuous transition--configuration C2 of \citet{Castillo}--and an acceleration well above the critical value ($\Gamma = 6.3, \Gamma_c \approx 5.1$), the effective surface tension was determined, $\gamma_{\rm 2D} = 2.9\pm0.1$~$\mu$N in two-dimensions. Using $H$ as the third dimension, \citet{Luu} obtained $\gamma \equiv \gamma_{\rm 2D}/H =1.5\pm0.1$~mN/m in 3D. Additionally, the mobility parameter $M$, which relates the interface speed with its curvature and surface tension, was measured by demonstrating the each mode obeys a Langevin-type equation. 
Finally, both $\gamma$ and $M$ are consistent with a simple order of magnitude estimation considering the characteristic energy, length, and time scales, which is very similar to what can be done for atomic systems.

\section{The liquid phase}
\label{s:liquidphase}

\subsection{Velocity distribution}
\label{subsec_veldist}

There is a wide range of parameters where the liquid phase is stable, which is simpler to characterize. Measurements of this state serve as tests for theories of confined granular matter. Using a high speed camera (normally operated using sequences of two rapid triggers) it is possible to measure the in-plane speed of all the particles. The resulting velocity distribution function $P(v)$ deviates strongly from a Maxwell--Boltzmann. This deviation is characterized by two properties. For small velocities the kurtosis $\kappa=\langle v^4\rangle/\langle v^2\rangle^2$, which equals 3 for a Maxwellian,  takes values larger than 3 for a broader distribution and smaller than three for a narrow one, being particularly sensitive to the range of small velocities. Experiments give $\kappa>3$ for all densities and particles used, with decreasing values for increasing $\Gamma$. The results differ, however, if the system has or has not a top lid where, in the later case, the confinement is produced as a balance between energy injection, dissipation and gravity.  In absence  of a top lid the kurtosis approaches 3 for increasing acceleration, therefore recovering a Maxwellian distribution. But when the lid is present, finite asymptotic values are obtained, different from 3 \citep{Olafsen1999,Urbach2001}.

\begin{figure*}[t!]
\begin{center}
\includegraphics[height=5.0cm]{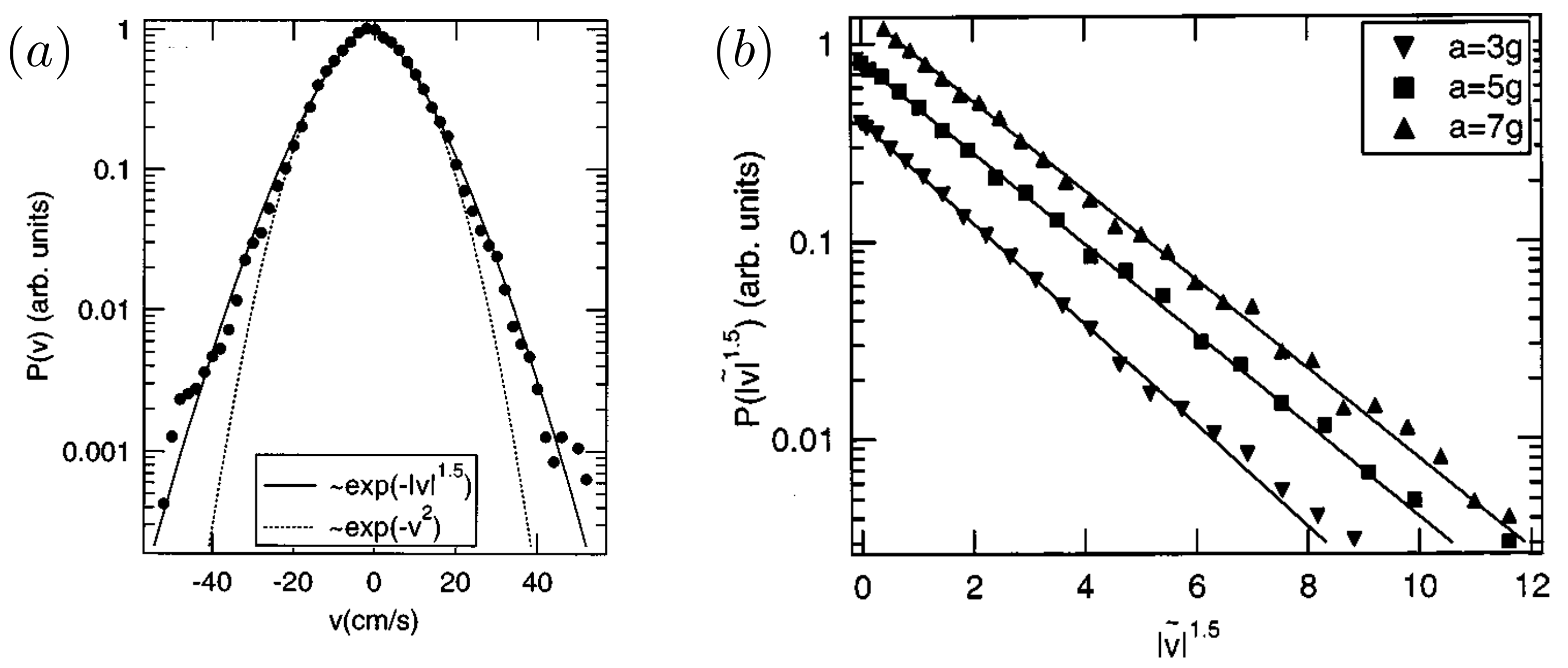} 
\caption{Probability distribution function (PDF) of the in-plane velocities $P(v)$ for a system with a top lid. (a) PDF in log-scale as a function of the velocity, where it is compared with a Maxwellian distribution. (b) PDF in log-scale as a function of $v^{1.5}$, where the linear trend is evident. Figure reprinted with permission from \citet{Losert1999b}. Copyright 1999, AIP Publishing LLC.}
\label{fig_Pv}
\end{center}
\noindent
\end{figure*}

For large velocities, the deviation from a Maxwellian distribution is characterized by the tails of $P(v)$ in log scale (shown in Fig.\ \ref{fig_Pv}). In systems with a top lid a slower than Maxwellian decay is observed, which is given roughly by $P(v)\sim e^{-v^{1.5}}$ \citep{Losert1999b,Urbach2001}. These overpopulated tails appear also in other configurations, as in free cooling granular gases.
The overpopulated tails disappear at high frequencies when the top lid is removed \citep{Olafsen1999,Urbach2001}. Together with the result for the kurtosis, the conclusion is that systems without lid at high frequencies are well described by a Maxwellian distribution.

The difference between experiments with and without top lid, indicates that the strong deviation from a Maxwell--Boltzmann distribution is related to the energy transfer mechanism from the vertical to the horizontal scale, where the anisotropy between the vertical and horizontal distributions becomes relevant. Under confinement, the vertical velocities are dominated by the walls and there is no thermalization (equipartition) between the two directions. Indeed,  the velocity distribution for the vertical direction, which can be measured in simulations, gives a vertical granular temperature that is much larger than the horizontal one \citep{rivas2}.

At the phase-separated regime, \citet{Olafsen} measured velocity distribution functions for the gas, collapsed and ordered states. The distributions collapse on a single curve when appropriately rescaled and show clear deviations from Maxwell statistics: a pronounced cusp is observed at low velocities and exponential tails at large velocities. The authors suggest that both types of deviations may be a consequence of clustering in momentum space, understood as the correlation between high density and low velocity when particles condense in real space. These dense regions in turn allow the other particles to access to larger velocities because of the reduction of the collisional frequency. 
Later, a more detailed study of the density and velocity statistics and correlations was reported \citep{Olafsen1999}. They demonstrated that the non-Gaussian nature of the velocity distribution increases as the acceleration is decreased from an homogeneous gas state ($\Gamma \approx 3$) down to the collapse transition line ($\Gamma \lesssim 1$), being almost Gaussian at  larger accelerations. It was also shown that velocity and density correlations become strong near this transition but for larger accelerations these quantities become uncorrelated. 

\subsection{Effective two-dimensional models}

Grains loose energy at collisions and the Q2D geometry allows granular matter to remain in fluidized states by the continuous transfer of energy from the top and bottom walls to the vertical degrees of freedom and, from these, to the in-plane horizontal degrees of freedom. Besides the direct simulation of three-dimensional systems that include naturally these processes, there has been an important effort in building effective two-dimensional models that include energy source terms.

The first model of this sort consists on adding random kicks to the grains, which are superimposed to the usual collisional dynamics \citep{Peng1998,VanNoije1998}. It is shown that, for all densities and inelasticities, the granular fluids adopts homogeneous steady states. Kinetic theory analysis of this model predicts that the distribution function is well approximated by a Maxwellian, except for an overpopulated tail $P(v)\sim \exp(-A v^{3/2})$
\citep{VanNoije1998}. The steady state presents  fluctuations that are of large amplitude and long correlation lengths, which is a result of the dynamics not conserving momentum \citep{VanNoije1999}. 

In the random kick model energy is injected to individual grains,  while in the Q2D geometry grains gain energy at collisions. 
As an improved approach to solve this problem, \citet{Barrat2001a} and \citet{Barrat2001b} introduced the random restitution coefficient model. Here grains move in two dimensions but the restitution coefficient that characterize collisions is a random variable that can adopt values larger than one, hence injecting energy. Different probability distributions for the restitution coefficient are used with the sole condition that the average injected energy vanishes to obtain steady states with finite energy. In all cases, the conclusions are qualitatively similar. Namely, that the velocity distribution function departs from a Maxwellian with a kurtosis larger than 3 and long tails for large velocities. The tails decay as $P(v)\sim e^{-A v^B}$, where the exponent $B$ depends on the probability distribution used for the restitution coefficient and lay in the range $[0.8-2]$.
The advantage of this model over the random kick description is that it correctly includes the energy injection in the horizontal dynamics at collision events.
However, in the long term, the total energy of the system performs a random walk because the inelastic hard sphere model with a random restitution coefficient does not present any characteristic intrinsic scale. 

Finally, \citet{Brito2013} introduced a model that solves the problems of the previous models to describe the in-plane dynamics in the Q2D geometry. Grains collide in two-dimensions and, besides the restitution coefficient $\alpha$ that is constant and smaller than one, a kick of a fixed speed $\Delta$ is given at the collision events, which are modeled by the collision rule
\begin{align}
{\bf v}_1^*={\bf v}_1 -\frac{1}{2} (1+\alpha) ({\bf v}_{12}\cdot\hat\sigma) \hat \sigma -\hat\sigma \Delta ,\\
{\bf v}_2^*={\bf v}_2 +\frac{1}{2} (1+\alpha) ({\bf v}_{12}\cdot\hat\sigma) \hat \sigma  +\hat\sigma \Delta ,
\end{align}
where ${\bf v}_{12}={\bf v}_1-{\bf v}_2$ is the relative velocity and $\hat\sigma$ points from particle 1 to 2. The system reaches stable steady states with a characteristic granular temperature that scales as $T\sim[\Delta/(1-\alpha^2)]^2$. Here, however, as a result of fixing an energy scale, the velocity distribution does not present long tails and the distribution resembles a Maxwellian. It could be possible that using a distribution for the values of $\Delta$ could change the situation.

\subsection{Hydrodynamics}

At the large, hydrodynamic scale the confined granular fluids present features characteristic of non-equilibrium systems. \citet{Prevost2002} measured the in-plane velocity--velocity correlations when the bottom layer is smooth or rough. It is found that for the smooth surface the correlations are long-ranged, decaying approximately  as $1/r^2$. When the rough surface is used the correlations decay exponentially with a correlation length that is larger for the longitudinal velocity (compression modes) compared with the transverse one (shear modes). The drastic difference between the two cases is due to the momentum dissipation in the later.

Using a rough bottom wall, which injects momentum randomly while dissipating it by friction, \citet{Gradenigo2011} and \citet{Puglisi2012}  studied experimentally and in simulations the role of friction in the development of velocity correlations in the hydrodynamic scale. The origin of the correlation is related to the absence of a fluctuation--dissipation relation between noise and friction. In all cases, correlations decay exponentially, and an algebraic decay is only obtained in absence of friction, consistent with the simulation of the random kick model \citep{VanNoije1999}.

Different hydrodynamic models have been formulated to describe the long-scale properties of confined systems. \citet{Khain2011} emphasized the relevance of the vertical motion, writing  separate equations for the vertical and horizontal temperatures. Energy transfer mechanisms were included which resulted in an effective negative compressibility for the in-plane pressure.
In a more general approach, \citet{Brito2013} consider a generic energy injection--dissipation  mechanism, which is coupled to standard hydrodynamic equations for granular fluids. Two regimes are found. For large dissipations or wavelengths the dynamics is dominated by dissipation and the longitudinal dynamics is governed by damped sound waves. In the opposite case, the diffusive energy transfer dominate the longitudinal  modes.

Finally, hydrodynamic equations can be obtained from effective two-dimensional models using the tools of kinetic theory. In particular, this program has been achieved in detail for the model presented at the end of the last section \citep{Brito2013,Brey2014,Brey2015,Soto2015}. The recently obtained equations have been completed with the corresponding transport coefficients. Adding the appropriate fluctuating terms these equations could be analyzed to compare with the experimental results. In the case of the solid-liquid transition, these equations must be coupled with one describing the bond-orientational order parameter evolution.

\section{Conclusions}
\label{s:concl}

Granular media in confined quasi two-dimensional geometry allows for a simultaneous study of the microscopic and collective dynamics, with the aim of finding models that describe the various phenomena that granular media present. We adopt the proposal that advances in the description of dense granular media could be attained by combining fluid dynamics with appropriate order parameters. In the confined geometry, grains are placed in a shallow box of only a few grains in height, while the lateral dimensions are large. The full box is vibrated periodically and the energy is injected by collisions of the grains with the walls, and later transferred by grain--grain collisions to the horizontal motion. This horizontal motion is monitored experimentally by optical means and is the main object of study.

For box heights of some tens of diameters, the vibrated granular fluid presents a liquid--vapor transition. The system is described by the integrated density over the vertical dimension as the order parameter. The average temperature is obtained from the balance between injection and dissipation and results to be a decreasing function of density. Consequently, the effective pressure presents regions of negative compressibility, which is the responsible of the phase separation. 

By decreasing the confining height, the granular layers show a liquid--solid transition. Depending on the vibration amplitude and frequency, the solid can occupy one or two layers, with different crystallographic symmetries. Qualitatively, the origin of the transition is similar to the previous case with negative compressibility being the responsible of the transition. However, to characterize it, it is found that besides density, the crystalline order should be used as order parameter. With this appropriate order parameter, the transition can be classified as being continuous or discontinuous. In the former case, critical exponents have been measured.

For both the liquid--vapor and liquid--solid transitions the phase separation is stable, with both phases coexisting. The interfaces fluctuate but are otherwise stable and the denser phase presents circular shapes, indicating that an effective surface tension exists. For the liquid--vapor transition, this effective tension was included in the model for the order parameter and, later, it was measured in simulations by using the Laplace law and the anisotropy of the stress tensor. For the liquid--solid transition, the spectral analysis of interface capillary waves allowed the experimental measurement of the effective surface tension.

In the homogeneous liquid phase, the in-plane velocity distribution functions deviate strongly from a Maxwellian. Long tails and kurtosis larger than for a Maxwellian are obtained in experiments. The origin of these deviations is the anisotropy between the vertical and horizontal motion. Effective two-dimensional models that incorporate the energy injection from the vertical to the horizontal motion describe some properties of the full three dimensional systems, particularly some features of the distribution function, but still the complete transfer mechanism are not captured. These effective models, because of their simplicity compared to the case of three dimensions, allows the construction of hydrodynamic equations that are coupled to different mechanisms of energy injection. To fully describe the phase transitions, the crystalline order parameter remains to be included.

\section*{Acknowledgements}

We acknowledge the support of Fondecyt Grants No. 1150393 (N.M.) and No. 1140778 (R.S.).

%
%


\begin{thebibliography}{40}

           \bibitem[{Amarouchene, Boudet, and Kellay(2008)}]{Amarouchene}  
\bibinfo{author}{Amarouchene Y, Boudet JF, and Kellay H}
  ({\bibinfo{year}{2008}}). 
  {\bibinfo{title}{Capillarylike Fluctuations at the Interface of Falling Granular Jets}}, 
  \bibinfo{journal}{Phys. Rev. Lett.} 
  {\bibinfo{volume}{100}}:
   \bibinfo{pages}{218001}.
   
         \bibitem[{Andreotti, Forterre and Pouliquen(2013)}]{AFP_Book}  
\bibinfo{author}{Andreotti B, Forterre Y and Pouliquen O}
  ({\bibinfo{year}{2013}}). 
  {\bibinfo{booktitle}{Granular media: Between fluid and solid}}, 
  \bibinfo{publisher}{(University Cambridge Press, Cambridge)} 
  

\bibitem[{Aranson et al.(2000)}]{Aranson2000}  
\bibinfo{author}{Aranson IS, Blair D, Kalatsky VA, Crabtree GW, Kwok WK, Vinokur VM and Welp U}
  ({\bibinfo{year}{2000}}). 
  {\bibinfo{title}{Electrostatically Driven Granular Media: Phase Transitions and Coarsening}}, 
  \bibinfo{journal}{Phys. Rev. Lett.} 
  {\bibinfo{volume}{84}}:
   \bibinfo{pages}{3306--3309}.


   
      \bibitem[{Aranson and Tsimring(2006)}]{RMP06}  
\bibinfo{author}{Aranson IS and Tsimring LS}
  ({\bibinfo{year}{2006}}). 
  {\bibinfo{title}{Patterns and collective behavior in granular media: Theoretical concepts}}, 
  \bibinfo{journal}{Rev. of Mod. Phys.} 
  {\bibinfo{volume}{78}}:
   \bibinfo{pages}{641--692}.
   
      \bibitem[{Aranson et al.(2008)}]{Aranson2008}  
\bibinfo{author}{Aranson IS, Snezhko A, Olafsen JS and Urbach JS}
  ({\bibinfo{year}{2008}}). 
  {\bibinfo{title}{Comment on "Long-Lived Giant Number Fluctuations in a Swarming Granular Nematic"}}, 
  \bibinfo{journal}{Science} 
  {\bibinfo{volume}{320}}:
   \bibinfo{pages}{612c--612c}.
   
      \bibitem[{Argentina, Clerc and Soto(2002)}]{Argentina}  
\bibinfo{author}{Argentina M, Clerc MG and Soto R}
  ({\bibinfo{year}{2002}}). 
  {\bibinfo{title}{van der Waals-like Transition in Fluidized Granular Matter}}, 
  \bibinfo{journal}{Phys. Rev. Lett.} 
  {\bibinfo{volume}{89}}:
   \bibinfo{pages}{044301}.
   
                  \bibitem[{Barrat and Trizac(2001)}]{Barrat2001a}  
\bibinfo{author}{Barrat A and Trizac E}
  ({\bibinfo{year}{2001}}). 
  {\bibinfo{title}{Inelastic hard spheres with random restitution coefficient: a new model for heated granular fluids}}, 
  \bibinfo{journal}{Advances in Complex Systems} 
  {\bibinfo{volume}{4}}:
   \bibinfo{pages}{299--307}.


               \bibitem[{Barrat, Trizac, and Fuchs(2001)}]{Barrat2001b}  
\bibinfo{author}{Barrat A, Trizac E and  Fuchs JN}
  ({\bibinfo{year}{2001}}). 
  {\bibinfo{title}{Heated granular fluids: The random restitution coefficient approach}}, 
  \bibinfo{journal}{Eur. Phys. J. E } 
  {\bibinfo{volume}{5}}:
   \bibinfo{pages}{161--170}.

 \bibitem[{Ben-Naim et al.(2001)}]{BenNaim2001}  
\bibinfo{author}{Ben-Naim E,  Daya ZA, Vorobieff P and  Ecke RE}
  ({\bibinfo{year}{2001}}). 
  {\bibinfo{title}{Knots and Random Walks in Vibrated Granular Chains}}, 
  \bibinfo{journal}{Phys. Rev. Lett.} 
  {\bibinfo{volume}{86}}:
   \bibinfo{pages}{1414--1417}.

 \bibitem[{Brey et al.(2014)}]{Brey2014}  
\bibinfo{author}{Brey JJ, Garc\'\i a de Soria MI, Maynar P and Buz\'on V}
  ({\bibinfo{year}{2014}}). 
  {\bibinfo{title}{Memory effects in the relaxation of a confined granular gas}}, 
  \bibinfo{journal}{Phys. Rev. E} 
  {\bibinfo{volume}{90}}:
   \bibinfo{pages}{032207}.
   
      \bibitem[{Brey et al.(2015)}]{Brey2015}  
\bibinfo{author}{Brey JJ, Buz\'on V, Maynar P and Garc\'\i a de Soria MI}
  ({\bibinfo{year}{2015}}). 
  {\bibinfo{title}{Hydrodynamics for a model of a confined quasi-two-dimensional granular gas}}, 
  \bibinfo{journal}{Phys. Rev. E} 
  {\bibinfo{volume}{91}}:
   \bibinfo{pages}{052201}.

                  \bibitem[{Brito, Risso and Soto(2013)}]{Brito2013}  
\bibinfo{author}{Brito R, Risso D and Soto R}
  ({\bibinfo{year}{2013}}). 
  {\bibinfo{title}{Hydrodynamic modes in a confined granular fluid}}, 
  \bibinfo{journal}{Phys. Rev. E} 
  {\bibinfo{volume}{87}}:
   \bibinfo{pages}{022209}.
   
            \bibitem[{Castillo, Mujica and Soto(2012)}]{Castillo}  
\bibinfo{author}{Castillo G, Mujica N and Soto R}
  ({\bibinfo{year}{2012}}). 
  {\bibinfo{title}{Fluctuations and Criticality of a Granular Solid-Liquid-like Phase Transition}}, 
  \bibinfo{journal}{Phys. Rev. Lett.} 
  {\bibinfo{volume}{109}}:
   \bibinfo{pages}{095701}.
   
   
               \bibitem[{Castillo, Mujica and Soto(2015)}]{CastilloPRE}  
\bibinfo{author}{Castillo G, Mujica N and Soto R}
  ({\bibinfo{year}{2015}}). 
  {\bibinfo{title}{Universality and criticality of a second-order granular solid-liquid-like phase transition}}, 
  \bibinfo{journal}{Phys. Rev. E} 
  {\bibinfo{volume}{91}}:
   \bibinfo{pages}{012141}.
   
      \bibitem[{Cartes, Clerc and Soto(2004)}]{Cartes}  
\bibinfo{author}{Cartes C, Clerc MG and Soto R}
  ({\bibinfo{year}{2004}}). 
  {\bibinfo{title}{van der Waals normal form for a one-dimensional hydrodynamic model}}, 
  \bibinfo{journal}{Phys. Rev. E} 
  {\bibinfo{volume}{70}}:
   \bibinfo{pages}{031302}.
   
   \bibitem[{Cheng et al.(2007)}]{Cheng2007}  
\bibinfo{author}{Cheng X, Varas G, Daniel C, Jaeger HM and Nagel SR}
  ({\bibinfo{year}{2007}}). 
  {\bibinfo{title}{Collective behavior in a granular jet: Emergence of a liquid with zero surface tension}}, 
  \bibinfo{journal}{Phys. Rev. Lett.} 
  {\bibinfo{volume}{99}}:
   \bibinfo{pages}{188001}.


\bibitem[{Cheng et al.(2008)}]{Cheng2008}  
\bibinfo{author}{Cheng X, Xu L, Patterson A, Jaeger HM and Nagel SR}
  ({\bibinfo{year}{2008}}). 
  {\bibinfo{title}{Towards the zero-surface-tension limit in granular fingering instability}}, 
  \bibinfo{journal}{Nature Physics} 
  {\bibinfo{volume}{4}}:
   \bibinfo{pages}{234--237}.

   
         \bibitem[{Clerc et al.(2008)}]{Clerc}  
\bibinfo{author}{Clerc MG, Cordero P, Dunstan J, Huff K, Mujica N, Risso D and Varas G}
  ({\bibinfo{year}{2008}}). 
  {\bibinfo{title}{Liquid-solid-like transition in quasi-one-dimensional driven granular media}}, 
  \bibinfo{journal}{Nature Physics} 
  {\bibinfo{volume}{4}}:
   \bibinfo{pages}{249--254}.

               \bibitem[{Clewett et al.(2012)}]{Clewett2012}  
\bibinfo{author}{Clewett JPD, Roeller K, Bowley RM, Herminghaus S and Swift MR}
  ({\bibinfo{year}{2012}}). 
  {\bibinfo{title}{Emergent surface tension in vibrated, noncohesive granular media}}, 
  \bibinfo{journal}{Phys. Rev. Lett.} 
  {\bibinfo{volume}{109}}:
   \bibinfo{pages}{228002}.

            \bibitem[{Duran(2001)}]{Duran}  
\bibinfo{author}{Duran J}
  ({\bibinfo{year}{2001}}). 
  {\bibinfo{title}{Rayleigh-Taylor instabilities in thin films of tapped powder}}, 
  \bibinfo{journal}{Phys. Rev. Lett.} 
  {\bibinfo{volume}{87}}:
   \bibinfo{pages}{254301}.
   
\bibitem[{G{\'e}minard and Laroche(2004)}]{Geminard2004}  
\bibinfo{author}{G{\'e}minard JC and Laroche C}
  ({\bibinfo{year}{2004}}). 
  {\bibinfo{title}{Pressure measurement in two-dimensional horizontal granular gases}}, 
  \bibinfo{journal}{Phys. Rev. E} 
  {\bibinfo{volume}{70}}:
   \bibinfo{pages}{021301}.

\bibitem[{Goldhirsch and Zanetti(1993)}]{Goldhirsch1993}  
\bibinfo{author}{Goldhirsch I and Zanetti G}
  ({\bibinfo{year}{1993}}). 
  {\bibinfo{title}{Clustering instability in dissipative gases}}, 
  \bibinfo{journal}{Phys. Rev. Lett.} 
  {\bibinfo{volume}{70}}:
   \bibinfo{pages}{1619--1622}.
   
      \bibitem[{Gradenigo et al.(2011)}]{Gradenigo2011}  
\bibinfo{author}{Gradenigo G, Sarracino A, Villamaina D and Puglisi A}
  ({\bibinfo{year}{2011}}). 
  {\bibinfo{title}{Non-equilibrium length in granular fluids: From experiment to fluctuating hydrodynamics}}, 
  \bibinfo{journal}{EPL} 
  {\bibinfo{volume}{96}}:
   \bibinfo{pages}{14004}.
   
               \bibitem[{Hohenberg and Halperin(1977)}]{hohenberg}  
\bibinfo{author}{Hohenberg PC and Halperin BI}
  ({\bibinfo{year}{1977}}). 
  {\bibinfo{title}{Theory of dynamic critical phenomena}}, 
  \bibinfo{journal}{Rev. Mod. Phys.} 
  {\bibinfo{volume}{49}}:
   \bibinfo{pages}{435--479}.
   

\bibitem[{Howell, Aronson and Crabtree(2001)}]{Howell2001}  
\bibinfo{author}{Howell DW, Aronson IS and Crabtree GW}
  ({\bibinfo{year}{2001}}). 
  {\bibinfo{title}{Dynamics of electrostatically driven granular media: Effects of humidity}}, 
  \bibinfo{journal}{Phys. Rev. E} 
  {\bibinfo{volume}{63}}:
   \bibinfo{pages}{050301}.

\bibitem[{Jeager, Nagel and Behringer(1996)}]{RMP96}  
\bibinfo{author}{Jaeger HM, Nagel SR and Behringer RP}
  ({\bibinfo{year}{1996}}). 
  {\bibinfo{title}{Granular solids, liquids, and gases}}, 
  \bibinfo{journal}{Rev. of Mod. Phys.} 
  {\bibinfo{volume}{68}}:
   \bibinfo{pages}{1259--1273}.
   
    \bibitem[{Khain and Aranson(2011)}]{Khain2011}  
\bibinfo{author}{Khain E and Aranson IS}
  ({\bibinfo{year}{2011}}). 
  {\bibinfo{title}{Hydrodynamics of a vibrated granular monolayer}}, 
  \bibinfo{journal}{Physical Review E} 
  {\bibinfo{volume}{84}}:
   \bibinfo{pages}{031308}.


    \bibitem[{Lobkovsky, Vega-Reyes and Urbach(2009)}]{Lobkovsky2009}  
\bibinfo{author}{Lobkovsky AE, Vega-Reyes F and Urbach JS}
  ({\bibinfo{year}{2009}}). 
  {\bibinfo{title}{The effects of forcing and dissipation on phase transitions in thin granular layers}}, 
  \bibinfo{journal}{Eur. Phys. J. Special Topics} 
  {\bibinfo{volume}{179}}:
   \bibinfo{pages}{113--122}.
   

\bibitem[{Losert, Cooper and Gollub(1999)}]{Losert1999}  
\bibinfo{author}{Losert W, Cooper DGW and Gollub JP}
  ({\bibinfo{year}{1999}}). 
  {\bibinfo{title}{Propagating front in an excited granular layer}}, 
  \bibinfo{journal}{Phys. Rev. E} 
  {\bibinfo{volume}{59}}:
   \bibinfo{pages}{5855--5861}.

   \bibitem[{Losert et al.(1999)}]{Losert1999b}  
\bibinfo{author}{Losert W, Cooper DGW, Delour J, Kudrolli A and Gollub JP }
  ({\bibinfo{year}{1999}}). 
  {\bibinfo{title}{Velocity statistics in excited granular media}}, 
  \bibinfo{journal}{Chaos} 
  {\bibinfo{volume}{9}}:
   \bibinfo{pages}{682--690}.


            \bibitem[{Luu et al.(2013)}]{Luu}  
\bibinfo{author}{Luu L-H, Castillo G, Mujica N and Soto R}
  ({\bibinfo{year}{2013}}). 
  {\bibinfo{title}{Capillarylike fluctuations of a solid-liquid interface in a noncohesive granular system}}, 
  \bibinfo{journal}{Phys. Rev. E} 
  {\bibinfo{volume}{87}}:
   \bibinfo{pages}{040202}.
   


   
                  \bibitem[{May et al.(2013)}]{May}  
\bibinfo{author}{May C, Wild M, Rehberg I and Huang K}
  ({\bibinfo{year}{2013}}). 
  {\bibinfo{title}{Analog of surface melting in a macroscopic nonequilibrium system}}, 
  \bibinfo{journal}{Phys. Rev. E} 
  {\bibinfo{volume}{88}}:
   \bibinfo{pages}{062201}.

      \bibitem[{Melby et al.(2005)}]{Urbach3}  
\bibinfo{author}{Melby P, Vega Reyes F, Prevost A, Robertson R, Kumar P, Egolf DA and Urbach JS}
  ({\bibinfo{year}{2005}}). 
  {\bibinfo{title}{The dynamics of thin vibrated granular layers}}, 
  \bibinfo{journal}{J. Phys. Cond. Mat.} 
  {\bibinfo{volume}{17}}:
   \bibinfo{pages}{S2689--S2704}.
   

   \bibitem[{Merminod, Berhanu and Falcon(2014)}]{Merminod2014}  
\bibinfo{author}{Merminod S, Berhanu M and Falcon E}
  ({\bibinfo{year}{2014}}). 
  {\bibinfo{title}{Transition from a dissipative to a quasi-elastic system of particles with tunable repulsive interactions}}, 
  \bibinfo{journal}{EPL} 
  {\bibinfo{volume}{106}}:
   \bibinfo{pages}{44005}.


\bibitem[{Narayan, Ramaswamy and Menon(2007)}]{Narayan2007}  
\bibinfo{author}{Narayan V, Ramaswamy S and Menon N}
  ({\bibinfo{year}{2007}}). 
  {\bibinfo{title}{Long-Lived Giant Number Fluctuations in a Swarming Granular Nematic}}, 
  \bibinfo{journal}{Science} 
  {\bibinfo{volume}{317}}:
   \bibinfo{pages}{105--108}.


   \bibitem[{Narayan, Ramaswamy and Menon(2008)}]{Narayan2008}  
\bibinfo{author}{Narayan V, Ramaswamy S and Menon N}
  ({\bibinfo{year}{2008}}). 
  {\bibinfo{title}{Response to Comment on "Long-Lived Giant Number Fluctuations in a Swarming Granular Nematic"}}, 
  \bibinfo{journal}{Science} 
  {\bibinfo{volume}{320}}:
   \bibinfo{pages}{612d--612d}.
              
      
            \bibitem[{N\'eel et al.(2014)}]{Neel}  
\bibinfo{author}{N\'eel B, Rondini I, Turzillo A, Mujica N and Soto R}
  ({\bibinfo{year}{2014}}). 
  {\bibinfo{title}{Dynamics of a first order transition to an absorbing state}}, 
  \bibinfo{journal}{Phys. Rev. E} 
  {\bibinfo{volume}{89}}:
   \bibinfo{pages}{042206}.

   \bibitem[{Olafsen and Urbach(1998)}]{Olafsen}  
\bibinfo{author}{Olafsen JS and Urbach JS}
  ({\bibinfo{year}{1998}}). 
  {\bibinfo{title}{Clustering, Order, and Collapse is a Driven Granular Monolayer}}, 
  \bibinfo{journal}{Phys. Rev. Lett.} 
  {\bibinfo{volume}{81}}:
   \bibinfo{pages}{4369--4372}.
   
   \bibitem[{Olafsen and Urbach(1999)}]{Olafsen1999}  
\bibinfo{author}{Olafsen JS and Urbach JS}
  ({\bibinfo{year}{1999}}). 
  {\bibinfo{title}{Velocity distributions and density fluctuations in a granular gas}}, 
  \bibinfo{journal}{Phys. Rev. E} 
  {\bibinfo{volume}{60}}:
   \bibinfo{pages}{R2468--R2471}.

   \bibitem[{Olafsen and Urbach(2001)}]{Urbach2001}  
\bibinfo{author}{Olafsen JS and Urbach JS}
  ({\bibinfo{year}{2001}}). 
  {\bibinfo{title}{Experimental Observations of Non-equilibrium Distributions and Transitions
in a 2D Granular Gas}}, 
  \bibinfo{booktitle}{Granular Gases},
  \bibinfo{editor}{P\"oschel T and Luding S (Eds)} 
   \bibinfo{publisher}{(Springer)} 
   \bibinfo{pages}{410--428}.
  
   
                     \bibitem[{Olafsen and Urbach(2005)}]{Olafsen2005}  
\bibinfo{author}{Olafsen JS and Urbach JS}
  ({\bibinfo{year}{2005}}). 
  {\bibinfo{title}{Two-dimensional melting far from equilibrium in a granular monolayer}}, 
  \bibinfo{journal}{Phys. Rev. Lett.} 
  {\bibinfo{volume}{95}}:
   \bibinfo{pages}{098002}.
   
                  \bibitem[{Orza et al.(1997)}]{Orza1997}  
\bibinfo{author}{Orza JAG, Brito R, Van Noije TPC and Ernst MH}
  ({\bibinfo{year}{1997}}). 
  {\bibinfo{title}{Patterns and long range correlations in idealized granular flows}}, 
  \bibinfo{journal}{International Journal of Modern Physics C} 
  {\bibinfo{volume}{8}}:
   \bibinfo{pages}{953--965}.
   
               \bibitem[{Oyarte et al.(2013)}]{Oyarte}  
\bibinfo{author}{Oyarte L, Guti\'errez P, Auma\^{i}tre S and Mujica N}
  ({\bibinfo{year}{2013}}). 
  {\bibinfo{title}{Phase Transition in a Out-of-equilibrium Monolayer of Dipolar Vibrated Grains}}, 
  \bibinfo{journal}{Phys. Rev. E} 
  {\bibinfo{volume}{87}}:
   \bibinfo{pages}{022204}.



               \bibitem[{Pacheco-V\'azquez, Caballero-Robledo and Ruiz-Su\'arez(2009)}]{Pacheco}  
\bibinfo{author}{Pacheco-V\'azquez F, Caballero-Robledo GA and Ruiz-Su\'arez JC}
  ({\bibinfo{year}{2009}}). 
  {\bibinfo{title}{Superheating in Granular Matter}}, 
  \bibinfo{journal}{Phys. Rev. Lett.} 
  {\bibinfo{volume}{102}}:
   \bibinfo{pages}{170601}.
   
      \bibitem[{Peng and Ohta(1998)}]{Peng1998}  
\bibinfo{author}{Peng G and Ohta T}
  ({\bibinfo{year}{1998}}). 
  {\bibinfo{title}{Steady state properties of a driven granular medium}}, 
  \bibinfo{journal}{Phys. Rev. E} 
  {\bibinfo{volume}{58}}:
   \bibinfo{pages}{4637}.

\bibitem[{P{\'e}rez-{\'A}ngel and Nahmad-Molinari(2011)}]{PerezAngel2011}  
\bibinfo{author}{P{\'e}rez-{\'A}ngel G and Nahmad-Molinari Y}
  ({\bibinfo{year}{2011}}). 
  {\bibinfo{title}{Bouncing, rolling, energy flows, and cluster formation in a two-dimensional vibrated granular gas}}, 
  \bibinfo{journal}{Phys. Rev. E} 
  {\bibinfo{volume}{84}}:
   \bibinfo{pages}{041303}.

      \bibitem[{Prevost et al.(2004)}]{Prevost}  
\bibinfo{author}{Prevost A, Melby P, Egolf DA and Urbach JS}
  ({\bibinfo{year}{2004}}). 
  {\bibinfo{title}{Non-equilibrium two-phase coexistence in a confined granular layer}}, 
  \bibinfo{journal}{Phys. Rev. E} 
  {\bibinfo{volume}{70}}:
   \bibinfo{pages}{050301(R)}.
   
      \bibitem[{Prevost, Egolf and Urbach(2002)}]{Prevost2002}  
\bibinfo{author}{Prevost A, Egolf DA and Urbach JS}
  ({\bibinfo{year}{2002}}). 
  {\bibinfo{title}{Forcing and Velocity Correlations in a Vibrated Granular Monolayer}}, 
  \bibinfo{journal}{Phys. Rev. Lett.} 
  {\bibinfo{volume}{89}}:
   \bibinfo{pages}{084301}.

      \bibitem[{Puglisi et al.(2012)}]{Puglisi2012}  
\bibinfo{author}{Puglisi A, Gnoli A, Gradenigo G, Sarracino A and Villamaina D}
  ({\bibinfo{year}{2012}}). 
  {\bibinfo{title}{Structure factors in granular experiments with homogeneous fluidization}}, 
  \bibinfo{journal}{J. Chem. Phys.} 
  {\bibinfo{volume}{136}}:
   \bibinfo{pages}{014704}.

               \bibitem[{Reis, Ingale and Shattuck(2006)}]{Reis2006}  
\bibinfo{author}{Reis P, Ingale RA and Shattuck MD}
  ({\bibinfo{year}{2006}}). 
  {\bibinfo{title}{Crystallization of a quasi-two-dimensional granular fluid}}, 
  \bibinfo{journal}{Phys. Rev. Lett.} 
  {\bibinfo{volume}{96}}:
   \bibinfo{pages}{258001}.


               \bibitem[{Reis, Ingale and Shattuck(2007)}]{Reis2007}  
\bibinfo{author}{Reis P, Ingale RA and Shattuck MD}
  ({\bibinfo{year}{2007}}). 
  {\bibinfo{title}{Caging dynamics in a granular fluid}}, 
  \bibinfo{journal}{Phys. Rev. Lett.} 
  {\bibinfo{volume}{98}}:
   \bibinfo{pages}{188301}.
   
   \bibitem[{Reyes and Urbach(2008)}]{Reyes2008}  
\bibinfo{author}{Reyes FV and Urbach JS}
  ({\bibinfo{year}{2008}}). 
  {\bibinfo{title}{Effect of inelasticity on the phase transitions of a thin vibrated granular layer}}, 
  \bibinfo{journal}{Phys. Rev. E} 
  {\bibinfo{volume}{78}}:
   \bibinfo{pages}{051301}.

                  \bibitem[{Rivas et al.(2011a)}]{rivas2}  
\bibinfo{author}{Rivas N, Cordero P, Risso D and Soto R}
  ({\bibinfo{year}{2011}}). 
  {\bibinfo{title}{Segregation in quasi two-dimensional granular systems}}, 
  \bibinfo{journal}{New Journal of Physics} 
  {\bibinfo{volume}{13}}:
   \bibinfo{pages}{055018}.     

      \bibitem[{Rivas et al.(2011b)}]{rivas}  
\bibinfo{author}{Rivas N, Ponce S, Gallet B, Risso D, Soto R, Cordero P and Mujica N}
  ({\bibinfo{year}{2011}}). 
  {\bibinfo{title}{Sudden chain energy transfer events in vibrated granular media}}, 
  \bibinfo{journal}{Phys. Rev. Lett.} 
  {\bibinfo{volume}{106}}:
   \bibinfo{pages}{088001}.
   
          
   
   \bibitem[{Rivas et al.(2012)}]{rivas3}  
\bibinfo{author}{Rivas N, Cordero P, Risso D and Soto R}
  ({\bibinfo{year}{2012}}). 
  {\bibinfo{title}{Characterization of the energy bursts in vibrated shallow granular systems}}, 
  \bibinfo{journal}{Granular Matter} 
  {\bibinfo{volume}{14}}:
   \bibinfo{pages}{157--162}.  
   
   
\bibitem[{Roeller et al.(2011)}]{Roeller2011}  
\bibinfo{author}{Roeller K, Clewett JPD, Bowley RM, Herminghaus S and Swift MR}
  ({\bibinfo{year}{2011}}). 
  {\bibinfo{title}{Liquid-gas phase Separation in confined vibrated dry granular matter}}, 
  \bibinfo{journal}{Phys. Rev. Lett.} 
  {\bibinfo{volume}{107}}:
   \bibinfo{pages}{048002}.
   
               \bibitem[{Royer et al.(2009)}]{Royer}  
\bibinfo{author}{Royer JR, Evans DJ, Oyarte L, Guo Q, Kapit E, M\"{o}bius ME, Waitukaitis SR and Jaeger HM}
  ({\bibinfo{year}{2009}}). 
  {\bibinfo{title}{High-speed tracking of rupture and clustering in freely falling granular streams}}, 
  \bibinfo{journal}{Nature} 
  {\bibinfo{volume}{459}}:
   \bibinfo{pages}{1110--1113}.
   
   
     \bibitem[{Safford et al.(2009)}]{Safford2009}  
\bibinfo{author}{Safford K, Kantor Y, Kardar M and Kudrolli A}
  ({\bibinfo{year}{2009}}). 
  {\bibinfo{title}{Structure and dynamics of vibrated granular chains: Comparison to equilibrium polymers}}, 
  \bibinfo{journal}{Phys. Rev. E} 
  {\bibinfo{volume}{79}}:
   \bibinfo{pages}{061304}.


     \bibitem[{Soto, Risso and Brito(2015)}]{Soto2015}  
\bibinfo{author}{Soto R, Risso D and Brito R}
  ({\bibinfo{year}{2015}}). 
  {\bibinfo{title}{Shear viscosity of a model for confined granular media}}, 
  \bibinfo{journal}{Phys. Rev. E} 
  {\bibinfo{volume}{90}}:
   \bibinfo{pages}{062204}.


\bibitem[{Ulrich and Zippelius(2012)}]{Ulrich2012}  
\bibinfo{author}{Ulrich S and Zippelius A}
  ({\bibinfo{year}{2012}}). 
  {\bibinfo{title}{Stability of freely falling granular streams}}, 
  \bibinfo{journal}{Phys. Rev. Lett.} 
  {\bibinfo{volume}{109}}:
   \bibinfo{pages}{166001}.




               \bibitem[{Van Noije and Ernst(1998)}]{VanNoije1998}  
\bibinfo{author}{Van Noije TPC  and Ernst MH}
  ({\bibinfo{year}{1998}}). 
  {\bibinfo{title}{Velocity distributions in homogeneous granular fluids: the free and the heated case}}, 
  \bibinfo{journal}{Gran. Matter} 
  {\bibinfo{volume}{1}}:
   \bibinfo{pages}{57--64}.


               \bibitem[{van Noije et al.(1999)}]{VanNoije1999}  
\bibinfo{author}{van Noije TPC, Ernst MH, Trizac E and Pagonabarraga I}
  ({\bibinfo{year}{1999}}). 
  {\bibinfo{title}{Randomly driven granular fluids: Large-scale structure}}, 
  \bibinfo{journal}{Phys. Rev. E} 
  {\bibinfo{volume}{59}}:
   \bibinfo{pages}{4236}.
   
   
      
%



\end{thebibliography}
\end{document}